\documentclass[aps,twocolumn,showpacs,pra,groupedaddress,superscriptaddress,nofootinbib,longbibliography]{revtex4-1}
\usepackage{graphicx,amsmath,amssymb,amsbsy,subfigure,hyperref,bbm,times}
\usepackage{color}

\providecommand{\openone}{\leavevmode\hbox{\small1\kern-3.8pt\normalsize1}}

\begin{document}

\title{Temperature effects on quantum non-Markovianity via collision models}

\author{Zhong-Xiao Man}
\email{zxman@mail.qfnu.edu.cn}
\affiliation{School of Physics and Physical Engineering, Shandong Provincial Key Laboratory of Laser Polarization and Information Technology, Qufu Normal University, 273165, Qufu, China}

\author{Yun-Jie Xia}
\email{yjxia@mail.qfnu.edu.cn}
\affiliation{School of Physics and Physical Engineering, Shandong Provincial Key Laboratory of Laser Polarization and Information Technology, Qufu Normal University, 273165, Qufu, China}

\author{Rosario Lo Franco}
\email{rosario.lofranco@unipa.it}
\affiliation{Dipartimento di Energia, Ingegneria dell'Informazione e Modelli Matematici, Universit\`{a} di Palermo, Viale delle Scienze, Edificio 9, 90128 Palermo, Italy}
\affiliation{Dipartimento di Fisica e Chimica, Universit\`a di Palermo, via Archirafi 36, 90123 Palermo, Italy}

\begin{abstract}
Quantum non-Markovianity represents memory during the system dynamics, which is typically weakened by the temperature.
We here study the effects of environmental temperature on the non-Markovianity of an open quantum system by virtue of collision models. The environment is simulated by a chain of ancillary qubits that are prepared in thermal states with a finite temperature $T$. Two distinct non-Markovian mechanisms are considered via two types of collision models, one where the system $S$ consecutively interacts with the ancillas and a second where $S$ collides only with an intermediate system $S'$ which in turn interacts with the ancillas. We show that in both models the relation between non-Markovianity and temperature is non-monotonic. In particular, revivals of non-Markovianity may occur as temperature increases. We find that the physical reason behind this behavior can be revealed by examining a peculiar system-environment coherence exchange, leading to ancillary qubit coherence larger than system coherence which triggers information backflow from the environment to the system. These results provide insights on the mechanisms underlying the counterintuitive phenomenon of temperature-enhanced quantum memory effects.
\end{abstract}


\maketitle

\section{Introduction}
In most practical situations a quantum system is open, being coupled to an environment that induces decoherence and dissipation of the system quantum properties \cite{open}. The dynamics of an open quantum system is usually described with a Markov approximation through a family of completely positive trace-preserving reduced dynamical maps and a corresponding quantum master equation with a Lindblad generator \cite{Lindblad1,Lindblad2}. In this case, the memoryless environment is assumed to be able to recover instantly from the interaction, which induces a monotonic one-way flow of information from the system to the environment.
However, due to the increasing capability to manipulate quantum systems, in many scenarios the Markov approximation is no longer valid leading to the occurrence of non-Markovian dynamics \cite{NM1,NM2} and a backflow of information from the environment to the system. The non-Markovian dynamics not only embodies an important physical phenomenon linked to dynamical memory effects but also proves useful to enhance practical procedures, such as quantum-state engineering and quantum control \cite{NMuse1,NMuse2,NMuse3,NMuse4,ManPRA2015,
LoFrancoNatCom,darrigo2012AOP,lofrancoPRB}.

Non-Markovianity has recently attracted considerable attention, particularly concerning the
formulation of its quantitative measures \cite{BLP,measure2,measure3,measure4,measure5,measure6,ACHL}, its experimental demonstration \cite{NMexp1,NMexp2,NMexp3,NMexp4,NMexp5} and the exploration of its origin \cite{origin1,origin2}. Nevertheless, the role of non-Markovianity for the assessment of the properties of non-equilibrium quantum systems has remained so far little explored \cite{NMentropy,cascaded,NMPower,MultiLandauer,Implications}.
Non-Markovian dynamics can lead to a new type of entropy production term which is indispensable to recover the fluctuation relations for entropy \cite{NMentropy}. In a bipartite system interacting dissipatively with a thermal reservoir in a cascaded model, the emerging non-Markovianity of one of the subsystems enables a heat flow with non-exponential time behavior \cite{cascaded}. By means of Landauer's principle, it has been also shown that memory effects are strategical in maintaining work extraction by erasure in realistic environments \cite{NMPower}. Moreover, non-Markovian dynamics can induce the breakdown of the validity of Landauer's principle \cite{MultiLandauer,Implications}.

An efficient tool that makes the study of quantum thermodynamics in the non-Markovian regime possible \cite{cascaded,MultiLandauer,Implications} is the \textit{collision} model \cite{Rau1963,colli1,colli2,colli3,colli4,colli5,colli6,colli7,colli8,colli9,colli10,colli11,
colli12,colli13,colli14,colli15,colli16,colli17,colli18,colli19,colli20,colli21,colli22,cicc2017}. In the collision model, the environment is taken as a collection of $N$ ancillas organized in a chain and the system of interest $S$ interacts, or collides, at each time step with an ancilla. It has been shown that when the ancillas are initially uncorrelated and no correlations are created among them along the process, a Lindblad master equation can be derived \cite{colli1,colli2}. By introducing either correlations in the initial state of the ancillas or inter-ancilla collisions, one can then recover the dynamics of any indivisible, and thus non-Markovian, channel \cite{colli3,colli4,colli5,colli6}.
In other words, the non-Makovian dynamics can be achieved in the collision model when the system-environment interaction is mediated by the ancillary degrees of freedom. In analogy to the well-known situation where the non-Markovian dynamics of a system arises when it is coherently coupled to an auxiliary system in contact with a Markovian bath, a class of Lindblad-type master equations for a bipartite system has been also found through collision models such that the reduced master equation of the system of interest is derived exactly \cite{colli7}. By constructing such composite collision models, one can simulate a lot of known instances of quantum non-Markovian dynamics, such as the emission of an atom into a reservoir with a Lorentzian, or multi-Lorentzian, spectral density or a qubit subject to random telegraph noise \cite{colli16}.

Albeit it is generally believed that quantum memory effects are more important at low temperatures \cite{Weiss}, the way temperature influences non-Markovianity depends on both  quantum thermodynamics and open quantum system dynamics. For a qubit subject to a dephasing bath with an Ohmic class spectrum, there exists a temperature-dependent critical value of the Ohmicity parameter for the onset of non-Markovianity which increases for high temperatures \cite{NMT1}. For a qubit in contact with a critical Ising spin thermal bath it has been then shown that the non-Markovianity decreases close to the critical point of the system in such a way that the higher the temperature, the higher the decrease \cite{NMT2}. Moreover, it is known that the non-Markovianity of a chromophore qubit in a super-Ohmic bath is reduced when the temperature increases \cite{NMT3}.
However, temperature may also enhance the non-Markovianity in some situations. For an inhomogeneous bosonic finite-chain environment, temperature has been shown to be a crucial factor in determining the character of the evolution and for certain parameter values non-Markovianity can increase with the temperature \cite{NMT4}. In a spin-boson model made of a two-level system which is linearly coupled to an environment of harmonic oscillators, a non-monotonic behavior of non-Markovianity as a function of temperature has been reported, with the system dynamics being strongly non-Markovian at low temperatures \cite{ClosPRA}. Another analysis, studying both entanglement and non-Markovianity measures to reveal how second-order weak-coupling master equations either overestimate or underestimate memory effects, suggests that non-Markonivity can be enriched by temperature \cite{NMT5}.

The above results, limited to specific situations, already show how subtle the effect of temperature on quantum non-Markovianity can be during an open system dynamics. In particular, the occurrence of temperature-enhanced memory effects remains counterintuitive and requires further studies which can unveil the underlying mechanisms. In this work we address this issue by means of suitable collision models, which reveal themselves specially advantageous to unveil the role of environmental elements in ruling the temperature-dependent non-Markovian dynamics of the system. We consider two types of collision models with different non-Markovian mechanisms, finding that in both models the variation of non-Markovianity as a function of temperature is not monotonic and providing the possible physical reason behind this phenomenon.

\section{Measure of non-Markovianity}
The degree of non-Markovianity in a dynamical process can be quantified by different measures, such as the BLP measure based on the distinguishability between the evolutions of two different initial states of the system \cite{BLP}, the LPP measure based on the volume of accessible states of the system \cite{measure2}, the RHP measure \cite{measure3} and the ACHL measure \cite{ACHL} based on the time-behavior of the master equation decay rates.

The trace distance between the evolutions of two different initial states $\rho_{1}(0)$ and $\rho _{2}(0)$ of an open system is one of the most employed quantifiers. Since a Markovian evolution can never increase the trace distance, when this happens it is a signature of non-Markovian dynamics of the system. Based on this concept, the non-Markovianity can be quantified by a the BLP measure $\mathcal{N}$ defined as
\cite{BLP}
\begin{equation}
\mathcal{N}=\max_{\rho _{1}(0),\rho _{2}(0)}\int_{\sigma >0}\sigma [t,\rho
_{1}(0),\rho _{2}(0)]dt,  \label{N}
\end{equation}
where $\sigma [t,\rho _{1}(0),\rho _{2}(0)]=dD[\rho _{1}(t),\rho_{2}(t)]/dt$ is the rate of change of the trace distance given by
\begin{equation}
D[\rho _{1}(t),\rho _{2}(t)]=\frac{1}{2}\mathrm{Tr}|\rho _{1}(t)-\rho
_{2}(t)|,  \label{Tra-Dis}
\end{equation}
with $|A|=\sqrt{A^{\dag }A}$.
To evaluate the non-Markovianity $\mathcal{N}$, one then has to find a specific pair of optimal initial states to maximize the time derivative of the trace distance. In Ref. \cite{optimal}, it is proved that the pair of optimal states is associated with two antipodal pure states on the surface of the Bloch sphere. We thus adopt, as usual, the pair of optimal initial states $\rho _{1,2}(0)=\left| \psi _{1,2}(0)\right\rangle\left\langle \psi _{1,2}(0)\right|$ with $\left| \psi
_{1,2}(0)\right\rangle =(\left| 0\right\rangle \pm \left| 1\right\rangle )/\sqrt{2}$.

Since the dynamics of the system in the collision model is implemented via $N$ equal discrete time steps, in the following the measure $\mathcal{N}$ shall be computed by substituting $\sigma [t,\rho _{1}(0),\rho _{2}(0)]dt $ with the difference
$\Delta D[n] = D[\rho_{1,n},\rho_{2,n}]-D[\rho_{1,n-1},\rho_{2,n-1}]$ between the trace distances at steps $n$ and $n-1$ and then summing up all the positive contributions, that is
\begin{equation}
\mathcal{N}=\max_{\rho _{1}(0),\rho _{2}(0)} \sum_{n,\Delta D[n]>0}^N \Delta D[n].
\quad (n=1,2,\ldots,N)
\end{equation}
The value of the final collision step $N$ is taken such as to cover all the oscillations of the trace distance during the evolution.

\section{Non-Markovianity in the direct collision model}\label{DCM}

\begin{figure}[tbp]
\begin{center}
{\includegraphics[width=\linewidth]{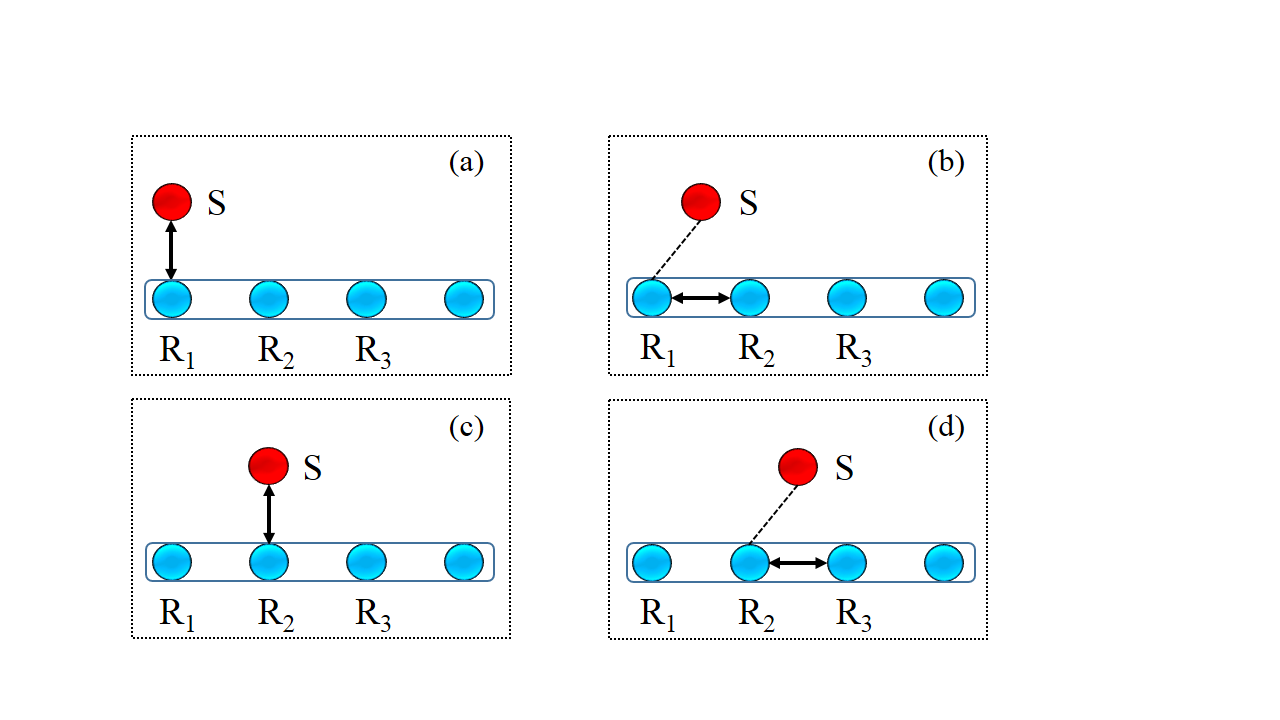}}
\end{center}
\caption{(Color online) Schematic diagram of the direct collision model. (a) The system $S$ collides with the ancilla qubit $R_{1}$ and they become correlated, as denoted by the dashed line in panel (b). (b) The intracollision between $R_{1}$ and $R_{2}$ takes place and the tripartite correlation $SR_{1}R_{2}$ may be generated. (c) The ancilla $R_{1}$ is traced out and the process is iterated: namely, the collision of $S$-$R_{2}$ in panel (c) is followed by that of $R_{2}$-$R_{3}$ in panel (d) and so on.}
\label{M1}
\end{figure}

In the first model, illustrated in Fig.~\ref{M1}, the system qubit $S$ directly interacts with the environment $\mathcal{R}$ which comprises $N$ identical qubits $R_{1},R_{2},\ldots,R_{N}$. The system qubit and a generic environment qubit are described, respectively, by the Hamiltonians ($\hbar=1$)
\begin{equation}
\hat{H}_{S}=\omega_{S} \hat{\sigma}_{z}^{S}/2,\quad
\hat{H}_{R}\equiv\hat{H}_{R_{n}}=\omega_{R} \hat{\sigma}_{z}^{R_{n}}/2,
\end{equation}
where $\hat{\sigma}_{z}^{\mu}=\left|1\right\rangle_{\mu}\left\langle1\right|-\left|0\right\rangle_{\mu}\left\langle0\right|$ is the Pauli operator and $\{\left|0\right\rangle_{\mu},\left|1\right\rangle_{\mu}\}$ are the logical states of the qubit $\mu=S,$ $R_{n}$ ($n=1,2,...,N$) with transition frequency $\omega_{\mu}$ (hereafter, for simplicity, we take $\omega_{R_{n}}=\omega_{R}=\omega_S = \omega$). The system-bath coupling is assumed to be ``white-noise'' (very large environment) so that the system never collides twice with the same qubit \cite{cicc2017}. As a consequence, at each collision step $n$ the system $S$ collides with a ``fresh" $R_{n}$.
Such a model can emulate, for a suitable combination of parameters and interactions, an atom coupled to a lossy cavity \cite{colli5}.

Among the possible choices for the interaction between $S$ and environment qubit $R_{n}$, here we focus on a Heisenberg-like coherent interaction described by the Hamiltonian
\begin{equation}\label{H}
\hat{H}_\mathrm{int}=g(\hat{\sigma}_{x}^{S}\otimes\hat{\sigma}_{x}^{R_{n}}+\hat{\sigma}_{y}^{S}\otimes\hat{\sigma}_{y}^{R_{n}}+\hat{\sigma}_{z}^{S}\otimes\hat{\sigma}_{z}^{R_{n}}),
\end{equation}
where $\hat{\sigma}_{j}^{\mu}$ ($j=x,y,z$) is the Pauli operator, $g$ denotes a coupling constant and each collision is described by a unitary operator $\hat{U}_{S,R_{n}}=e^{-i\hat{H}_\mathrm{int}\tau}$, $\tau$ being the collision time.
By means of the equality
\begin{equation}\label{eq}
e^{i\frac{\phi}{2}(\hat{\sigma}_{x}\otimes\hat{\sigma}_{x}+\hat{\sigma}_{y}\otimes\hat{\sigma}_{y}+\hat{\sigma}_{z}\otimes\hat{\sigma}_{z})}
=e^{-i\frac{\phi}{2}}(\cos\phi \ \hat{\mathbb{I}} +i\sin\phi \ \hat{\mathcal{S}})
\end{equation}
with $\hat{\mathbb{I}}$ the identity operator and $\hat{\mathcal{S}}$ the two qubit swap operator with the action  $\left|\psi_{1}\right\rangle\otimes\left|\psi_{2}\right\rangle\rightarrow\left|\psi_{2}\right\rangle\otimes\left|\psi_{1}\right\rangle$
for all $\left|\psi_{1}\right\rangle,\left|\psi_{2}\right\rangle\in \mathbb{C}^{2}$, the unitary time evolution operator can be written as
\begin{equation}\label{swapSR1}
\hat{U}_{SR_{n}}=(\cos J)\ \hat{\mathbb{I}}_{SR_{n}}+i(\sin J)\ \hat{\mathcal{S}}_{SR_{n}},
\end{equation}
where $J=2g\tau$ is a dimensionless interaction strength between $S$ and $R_{n}$ which is supposed to be the same for any $n=1,2,\ldots,N$.
It is immediate to see that $J=\pi/2$ induces a complete swap between the state of $S$ and that of $R_{n}$. Thus, $0<J<\pi/2$ means a partial swap conveying the intuitive idea that, at each collision, part of the information contained in the state of $S$ is transferred into $R_{n}$.
In the ordered basis $\{\left|00\right\rangle_{SR_{n}},\left|01\right\rangle_{SR_{n}},\left|10\right\rangle_{SR_{n}},\left|11\right\rangle_{SR_{n}}\}$, $\hat{U}_{SR_{n}}$ reads
\begin{equation}\label{swapSR2}
\hat{U}_{SR_{n}}=\left(
                   \begin{array}{cccc}
                     e^{iJ} & 0 & 0 & 0 \\
                     0 & \cos J & i\sin J & 0 \\
                     0 & i\sin J & \cos J & 0 \\
                     0 & 0 & 0 & e^{iJ} \\
                   \end{array}
                 \right).
\end{equation}
In the present model, the non-Markovian dynamics of the system is introduced via the interactions between two nearest-neighbor qubits $R_{n}$ and $R_{n+1}$. Such interactions are described by an operation similar to that of Eq.~(\ref{swapSR2}), namely
\begin{equation}\label{swapRR}
\hat{V}_{R_{n}R_{n+1}}=\left(
                   \begin{array}{cccc}
                     e^{i\Omega} & 0 & 0 & 0 \\
                     0 & \cos \Omega & i\sin \Omega & 0 \\
                     0 & i\sin \Omega & \cos \Omega & 0 \\
                     0 & 0 & 0 & e^{i\Omega} \\
                   \end{array}
                 \right),
\end{equation}
where $0\leq\Omega\leq\pi/2$ is the dimensionless $R_{n}$-$R_{n+1}$ interaction strength which is taken to be the same for any $n$.

\begin{figure}[tbp]
\begin{center}
{\includegraphics[width=0.9\linewidth]{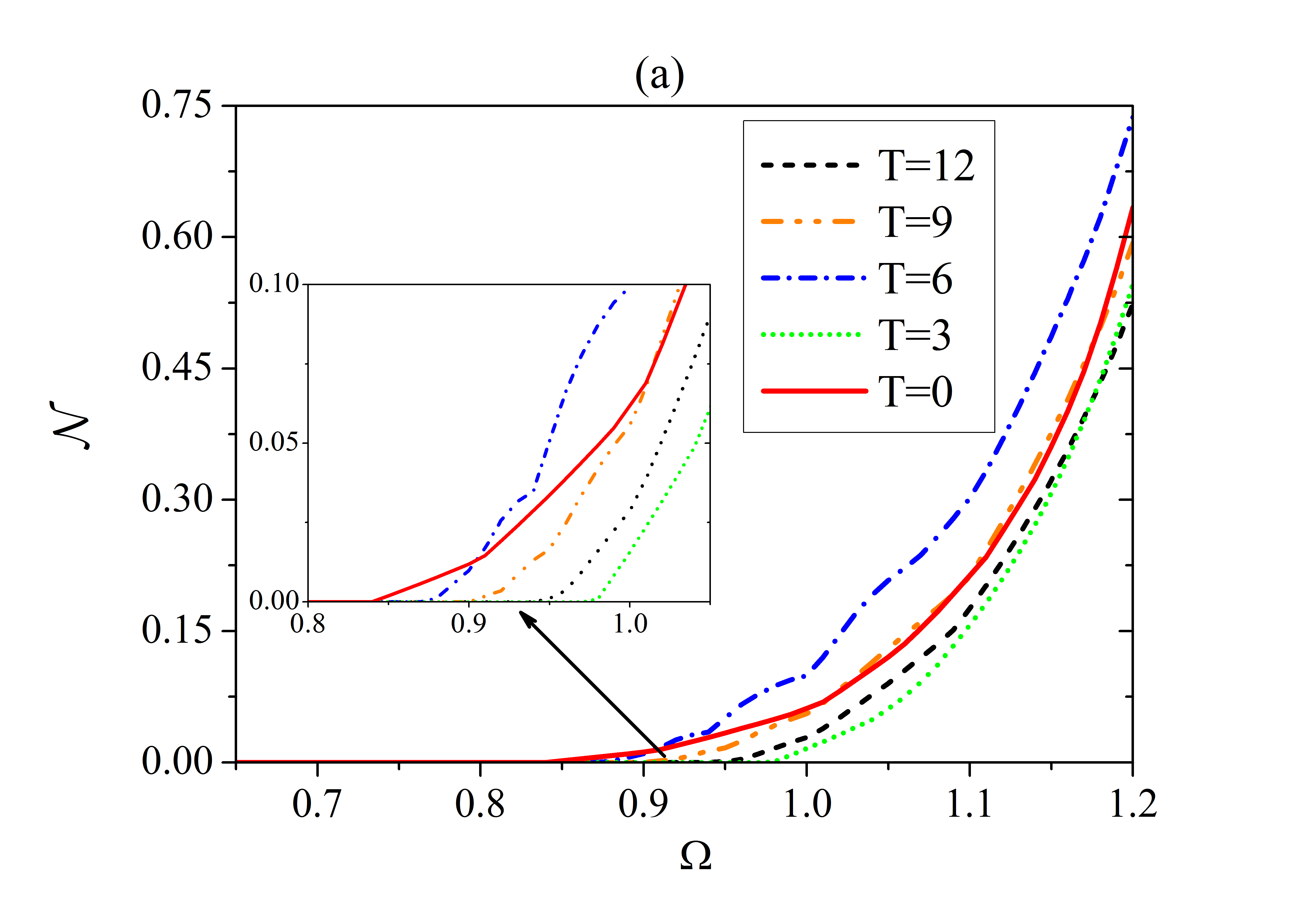} %
\includegraphics[width=0.9\linewidth]{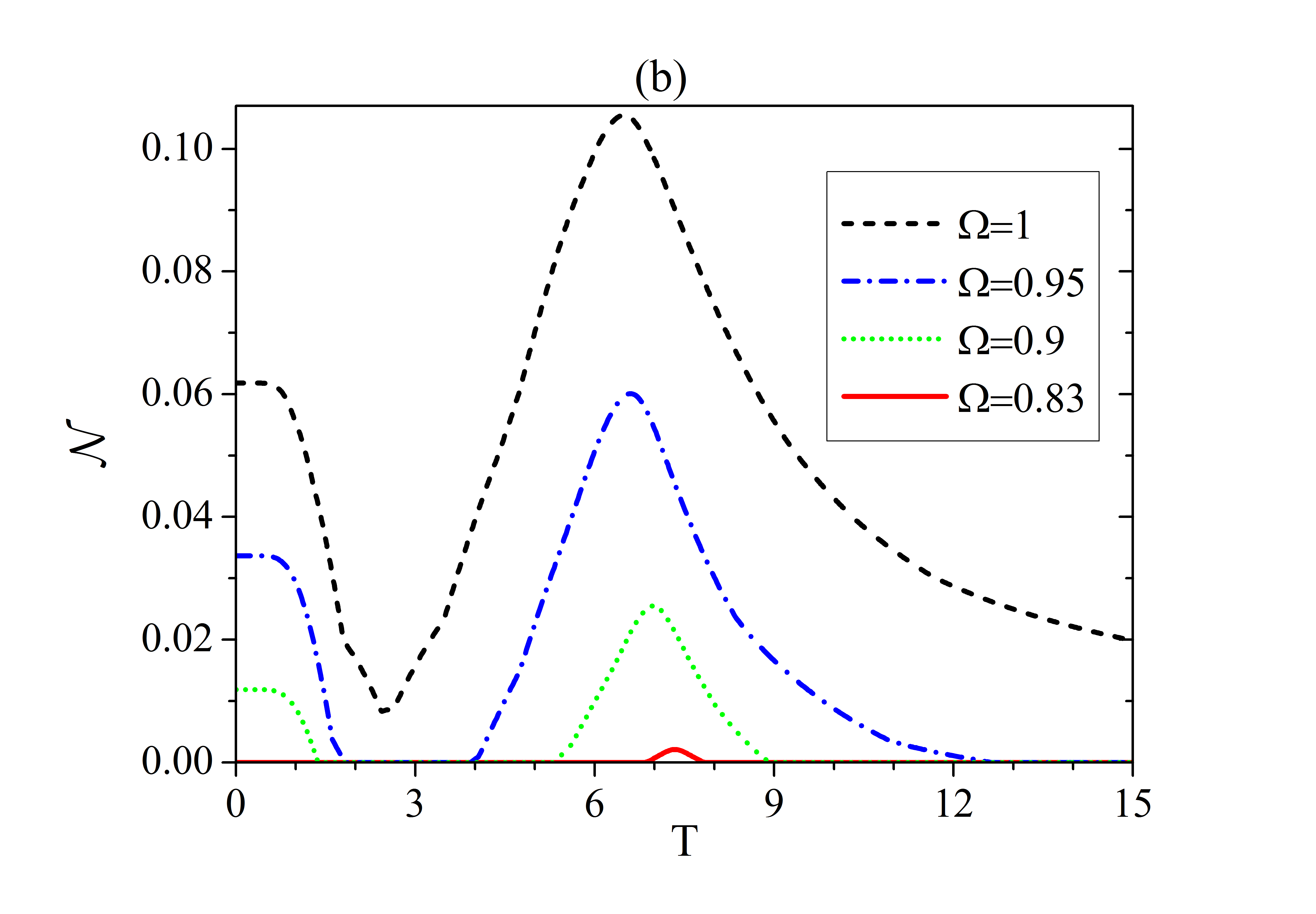} %
}
\end{center}
\caption{(Color online) \textbf{(a)} Non-Markovianity $\mathcal{N}$ vs. the collision strength $\Omega$ between the environment qubits for different temperatures $T$ of the environment. \textbf{(b)} Non-Markovianity $\mathcal{N}$ vs. the temperature $T$ for different $\Omega$. In both plots, the remaining parameters are given by $\omega=5\omega_0$ and $J=0.3=0.06\omega/\omega_0$ (small collision time and weak interaction between system and environment qubits).}
\label{NM-Omiga-T}
\end{figure}

As illustrated in Fig.~\ref{M1} exemplifying the first two steps of collisions, in each step we consider the ordered triplet $(S, R_{n-1}, R_{n})$ in such a way that after the collision between $S$ and $R_{n-1}$ via the unitary operation $\hat{U}_{SR_{n-1}}$, the system shifts by
one site while $R_{n-1}$ collides with $R_{n}$ via $\hat{V}_{R_{n-1}R_{n}}$. Notice that $R_{n-1}$-$R_n$ collision occurs before $S$-$R_n$ collision so that $S$ and $R_n$ are already correlated before they collide with each other. The three qubits after the two collisions can now be all correlated with the total state $\rho_{SR_{n-1}R_{n}}$ (the correlations are labeled by the dashed lines in Fig.~\ref{M1}). Then, we trace out the qubit $R_{n-1}$ giving rise to the reduced state $\rho_{SR_{n}}$ of $S-R_{n}$ and proceed to the next step with the new ordered triplet $(S, R_{n}, R_{n+1})$.
Under the actions of $\hat{U}_{SR_{n}}$ of Eq.~(\ref{swapSR2}) and $\hat{V}_{R_{n}R_{n+1}}$ of Eq.~(\ref{swapRR}), the total state of $SR_{n}R_{n+1}$ at the step $n$ is obtained from the step $n-1$ as
\begin{eqnarray}\label{state-n}
&\rho_{SR_{n}R_{n+1}}=&\nonumber\\
&\hat{V}_{R_{n}R_{n+1}}\hat{U}_{SR_{n}}\left(\rho_{SR_{n}}\otimes
\rho_{R_{n+1}}\right)\hat{U}_{SR_{n}}^{\dag} \hat{V}_{R_{n}R_{n+1}}^{\dag},&
\end{eqnarray}
where $\rho_{R_{n+1}}\equiv\rho_{R}$ is the pre-collision state of the environmental qubit.
Here, to reveal the effect of environmental temperature on the non-Markovianity, we assume the environmental qubits are initially prepared in the same thermal states $\rho_{R}=e^{-\beta \hat{H}_{R}}/Z$ at temperature $T_R$, where $\beta=1/k_B T_R$ ($k_B$ being the Boltzmann constant) and $Z$ is the partition function. In our analysis, we consider a dimensionless temperature $T$ defined by $T \equiv k_BT_R / (\hbar\omega_0)$, where $\omega_0$ is a reference frequency. We also take values of $J\ll\omega/\omega_0$ so to have small collision times and a weak interaction between the system and the environment qubits.

\begin{figure}[tbp]
\begin{center}
{\includegraphics[width=\linewidth]{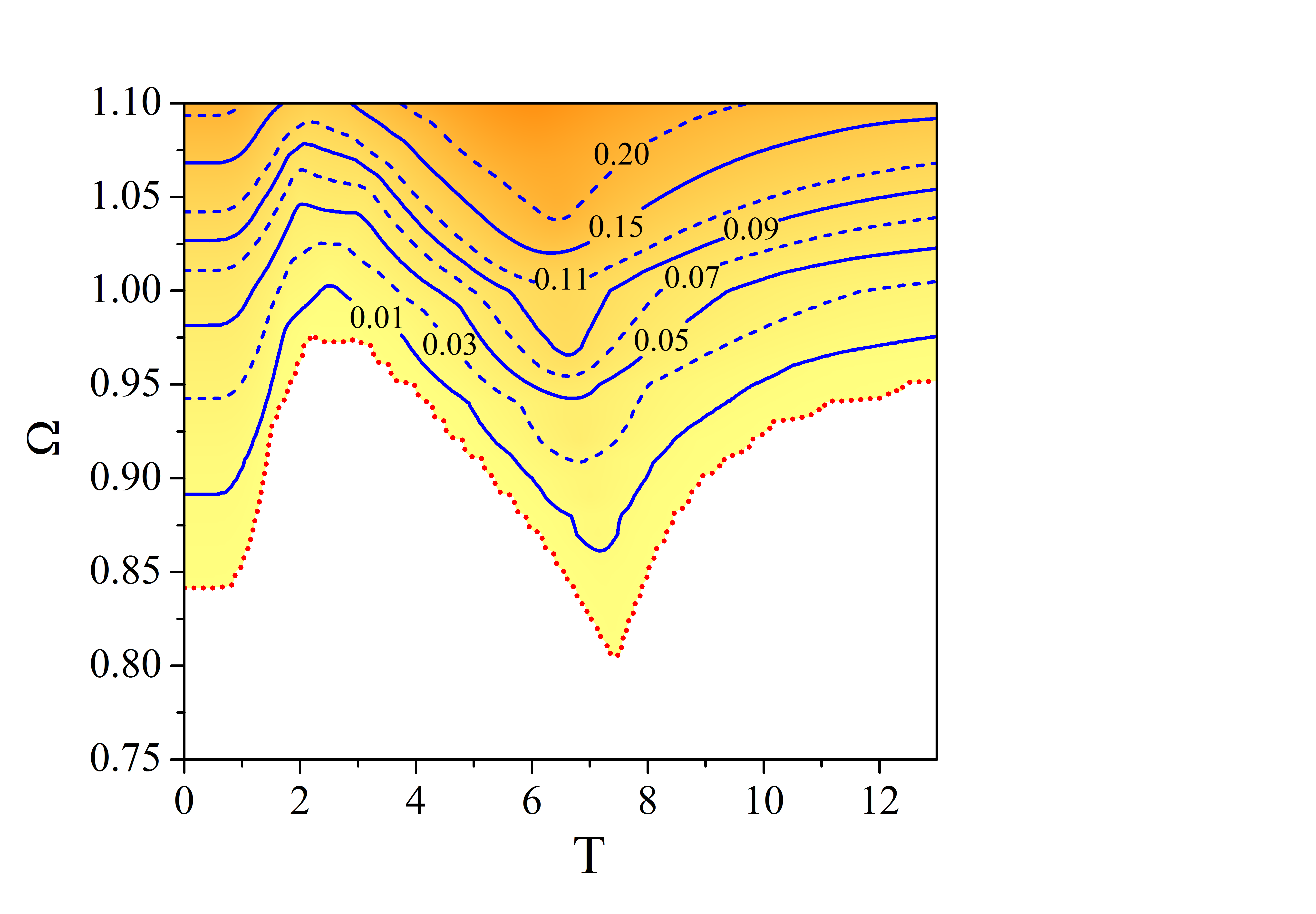} %
}
\end{center}
\caption{(Color online) Contour plot of the non-Markovianity $\mathcal{N}$ for different $T$ and $\Omega$.
The non-Markovian regime is colored while the Markovian regime is white. The blue lines are the contour lines of $\mathcal{N}$. The
dotted red line is the curve of the thresholds of $\Omega$ triggering non-Markovian dynamics.
The remaining parameters are given by $\omega=5\omega_0$ and $J=0.3=0.06\omega/\omega_0$ (small collision time and weak interaction between system and environment qubits).}
\label{cont-NM}
\end{figure}

In such a model, the system experiences a homogenization process and reaches asymptotically the very same state $\rho_{R}$ \cite{Implications}. The forward transfer of the lost information of the system $S$ via intracollisions of environment qubits triggers dynamical memory effects of the system, so that the non-Markovianity is closely related to the intracollision strength $\Omega$. Fig.~\ref{NM-Omiga-T}(a) shows the dependence of non-Markovianity $\mathcal{N}$ on $\Omega$ for different temperatures $T$ of the environment. In both zero temperature ($T=0$) and thermal environments ($T>0$), the non-Markovianity is activated when $\Omega$ exceeds a given threshold (see the inset of Fig.~\ref{NM-Omiga-T}(a) for a more evident demonstration) and then monotonically increases with $\Omega$. From this first analysis, it emerges that the thermal environment does not affect the monotonic relation between $\mathcal{N}$ and $\Omega$, while the thresholds of $\Omega$ triggering the non-Markovianity depend on the temperature.
On the other hand, the variations of non-Markovianity $\mathcal{N}$ with respect to the temperature $T$ can be rich and non-monotonic, as shown in Fig.~\ref{NM-Omiga-T}(b).
For relatively small values of $\Omega$ (e.g., $\Omega=0.83$), the increase of $T$ can enable the non-Markovianity which maintains nonzero values within a finite region of $T>0$. For larger values of $\Omega$, the system dynamics exhibits non-Markovian character already for a zero-temperature environment. In this case, the non-Markovianity approximately exhibits a plateau for small $T$ and then experiences successive decreasing and increasing behaviors, eventually vanishing at high temperatures. For particular values of the environment qubits interaction strength (e.g., $\Omega=0.9$, $0.95$), when $T$ increases we also observe that the non-Markovianity $\mathcal{N}$ may vanish within a finite interval of $T$ and then revive again. In other words, manipulations of the environment temperature $T$ can induce successive transitions between non-Markovian and Markovian regimes for the system dynamics.
A comprehensive picture for the dependence of $\mathcal{N}$ on $T$ and $\Omega$ is shown in Fig.~\ref{cont-NM}, where we can see the non-Markovianity thresholds of $\Omega$ (identified by the dotted red line) for a given $T$ and the crossovers between non-Markovian and Markovian regimes as $T$ increases for a given $\Omega$.

\begin{figure}[tbp]
\begin{center}
{\includegraphics[width=\linewidth]{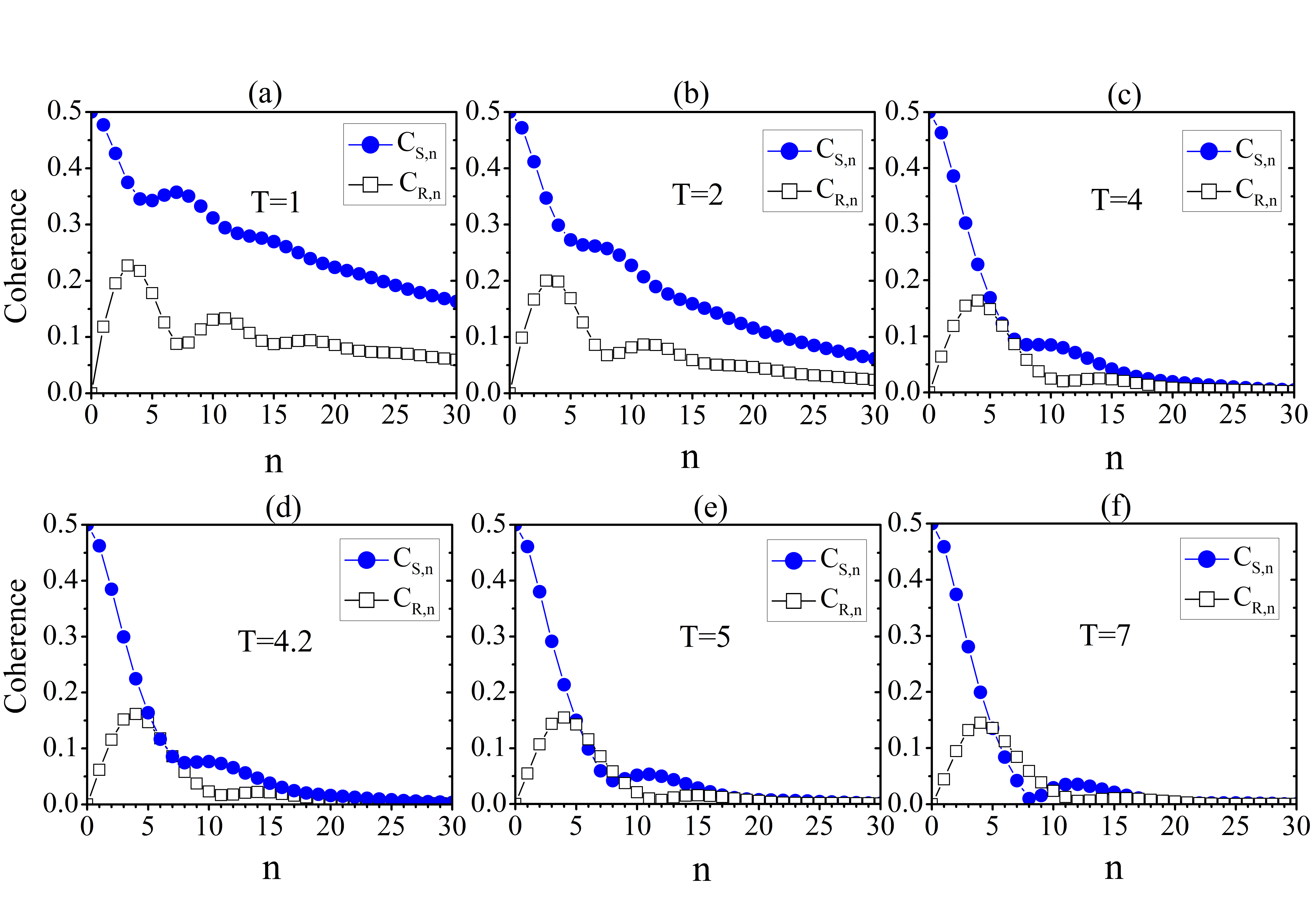}
}
\end{center}
\caption{(Color online) Coherences $C_{S,n}$ of the system $S$ and $C_{R,n}$ of the environment qubit $R_{n+1}$ after the $n$-th collision as a function of $n$.
The relevant parameters are $\Omega=0.95$, $\omega=5\omega_0$ and $J=0.3=0.06\omega/\omega_0$ (small collision time and weak interaction between system and environment qubits).}
\label{M1coh}
\end{figure}

In order to gain a deeper understanding of the temperature effects on the non-Markovianity, we examine the coherence of the system, which is related to the trace distance of this model as $D[\rho_{1,n},\rho_{2,n}]=2C_{S,n}$, where $C_{S,n}=|\left\langle0\right|\rho_{S,n}\left|1\right\rangle|=|\left\langle1\right|\rho_{S,n}\left|0\right\rangle|$ is the coherence degree of the state $\rho_{S,n}$ of $S$ after the $n$-th collision, the initial state being prepared in
$\left| \psi\right\rangle_{S,0} =(\left| 0\right\rangle_{S,0} \pm \left| 1\right\rangle_{S,0} )/\sqrt{2}$. Notice that the latter is a \textit{bona fide} quantifier of coherence, being the half of the so-called $l_1$-norm measure within a resource theory \cite{baumgratzPRL,streltsovRMP}.  Therefore, the initial coherence of $S$ has the maximum value $C_{S,0}=0.5$. The temporary growth of $C_{S,n}$ thus serves as a witness for the onset of non-Markovian dynamics. Moreover, to assess the role of the environmental constituents, we consider the coherence $C_{R,n}$ of the environment qubit $R_{n+1}$ transferred from $S$ after the $n$-th collision of $S$-$R_{n}$-$R_{n+1}$.
In Fig.~\ref{M1coh}(a)-(f), we illustrate the evolution of $C_{S,n}$ and $C_{R,n}$ versus $n$ for different temperatures with $\Omega=0.95$, whose non-Markovian character is plotted in Fig.~\ref{NM-Omiga-T}(b) (blue dot-dashed curve). An overall comparison of the panels in Fig.~\ref{M1coh}(a)-(f), which indicate a temperature range from $T=1$ to $T=7$, verifies the fact that the initial increase of temperature speeds up the decay of the system coherence $C_{S,n}$. Therefore, on the one hand, the increase of temperature suppresses and eventually terminates the non-Markovianity, as seen in Fig.~\ref{M1coh}(a)-(b) for $T=1$ and $T=2$ and already confirmed in Fig.~\ref{NM-Omiga-T}(b). On the other hand, however, the quick decay of $C_{S,n}$ can cause the coherence $C_{R,n}$ of the environment qubit $R_{n+1}$ to approach (see Fig.~\ref{M1coh}(c)) and even to exceed (see Fig.~\ref{M1coh}(d)-(e)-(f)) the coherence $C_{S,n}$ of the system. This behavior in turn induces the information backflow from the environment to the system, namely, a revival of non-Markovian regime. In fact, $C_{R,n}$ overcomes $C_{S,n}$ in correspondence to the recovery of the non-Markovian character of the system dynamics from a Markovian one (compare Fig.~\ref{M1coh}(d)-(e)-(f) and the blue dot-dashed curve of Fig.~\ref{NM-Omiga-T}(b)).
In the high temperature regime, the system coherence decays more quickly and the non-Markovian dynamics will cease if the intracollision strength $\Omega$ is not sufficiently large.  For instance, from Fig.~\ref{NM-Omiga-T}(b) one sees that at $T=10$ the non-Markovianity vanishes if $\Omega=0.9$, while it remains nonzero when $\Omega=0.95$. In other words, to get non-Markovian dynamics (quantum memory effects) at high temperatures, one has to increase the interaction strength $\Omega$ between environmental ancillary qubits, which allows a more efficient transfer of environmental quantum coherence

\section{Non-Markovianity in the indirect collision model}

\begin{figure}[tbp]
\begin{center}
{\includegraphics[width=\linewidth]{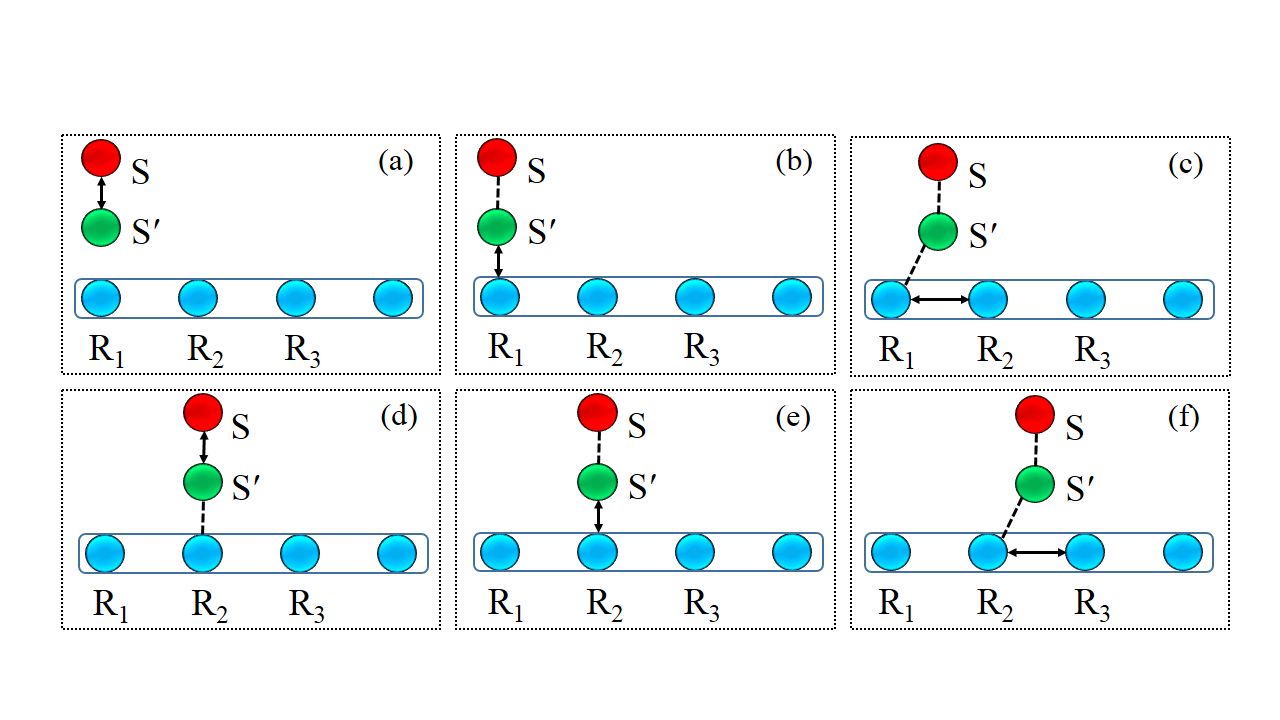}}
\end{center}
\caption{(Color online) Schematic diagram of the indirect collision model. (a) The system $S$ collides with the intermediate qubit $S'$ and they become correlated, as denoted by the dashed line in panel (b). (b) The qubit $S'$ interacts with $R_{1}$ and the correlation among $SS'R_{1}$ is generated, as denoted by the dashed line in (c). (c) The intracollision between $R_{1}$ and $R_{2}$ takes place and the correlation of $SS'R_{1}R_{2}$ is then established. (d) The ancilla qubit $R_{1}$ is traced out and the process is iterated: namely, the collision of $S$-$S'$ in (d) is  followed by the collision $S'$-$R_{2}$ in (e) and $R_{2}$-$R_{3}$ in (f) and so on.}
\label{M2}
\end{figure}

We now consider a mechanism of non-Markovian dynamics based on another collision model, where the interaction of the system qubit $S$ with the environment qubit $R_{n}$ is mediated by an intermediate qubit $S'$, as depicted in Fig.~\ref{M2}. Such a scenario implies that the information contained in $S$ is first transferred to $S'$ and then damped into $\mathcal{R}$ via the collisions between $S'$ and $R_{n}$. It is known that, in the absence of environmental intracollisions, this composite model can emulate (for short collision times and Jaynes-Cummings-type interactions) a two-level atom in a lossy cavity, $S'$ playing the role of the cavity mode \cite{colli16}. For straightforward extension, in the presence of environmental intracollisions, this model may represent a two-level atom in a reservoir with a photonic band gap \cite{colli16,laurapseudo,ManSciRep2015}.

We choose the Heisenberg-type coherent interaction between $S$ and $S'$, with interaction strength $0\leq \kappa\leq\pi/2$, represented by the unitary operator $\hat{U}_{SS'}$, analogous to that of Eq.~(\ref{swapSR2}), having the form
\begin{equation}\label{swapSS}
\hat{U}_{SS'}=\left(
                   \begin{array}{cccc}
                     e^{i\kappa} & 0 & 0 & 0 \\
                     0 & \cos \kappa & i\sin \kappa & 0 \\
                     0 & i\sin \kappa & \cos \kappa & 0 \\
                     0 & 0 & 0 & e^{i\kappa} \\
                   \end{array}
                 \right).
\end{equation}
The unitary operators $\hat{U}_{S'R_{n}}$ and $\hat{V}_{R_{n}R_{n+1}}$ representing, respectively, the $S'$-$R_{n}$ interaction and the interaction between adjacent environment qubits are the same of Eqs.~(\ref{swapSR2}) and (\ref{swapRR}) with interaction strengths $J$ and $\Omega$.

As shown in Fig.~\ref{M2}, in each round of collisions we deal with four qubits $(S, S', R_{n-1}, R_{n})$ in such a way that, after the collisions of $S$-$S'$ and $S'$-$R_{n-1}$, the qubits $S$ and $S'$ shift by one site while $R_{n-1}$ collides with $R_{n}$, which results in the correlated total state $\rho_{SS'R_{n-1}R_{n}}$ (the correlations are indicated by dashed lines). Then, we trace out the qubit $R_{n-1}$ obtaining the reduced state $\rho_{SS'R_{n}}$ of $SS'R_{n}$ and proceed to the next step with the new ordered group $(S, S', R_{n}, R_{n+1})$.
As a consequence, the total state of $SS'R_{n}R_{n+1}$ at the $n$-th collision is determined from the $(n-1)$-th collision as
\begin{eqnarray}\label{state-SSRR}
&\rho_{SS'R_{n}R_{n+1}}=&\nonumber\\
&\hat{V}_{R_{n}R_{n+1}}\hat{U}_{S'R_{n}}\hat{U}_{SS'}\left(\rho_{SS'R_{n}}\otimes \rho_{R_{n+1}}\right)\hat{U}_{SS'}^{\dag}\hat{U}_{S'R_{n}}^{\dag} \hat{V}_{R_{n}R_{n+1}}^{\dag}.&\nonumber\\
\end{eqnarray}
The temperature effects are included in this model by considering the qubit $S'$ and all the environmental qubits prepared in the same thermal states $\rho_{R}=e^{-\beta \hat{H}_{R}}/Z$ with temperature $T$.

\subsection{Absence of collisions between environment qubits}

\begin{figure}[tbp]
\begin{center}
{\includegraphics[width=0.9\linewidth]{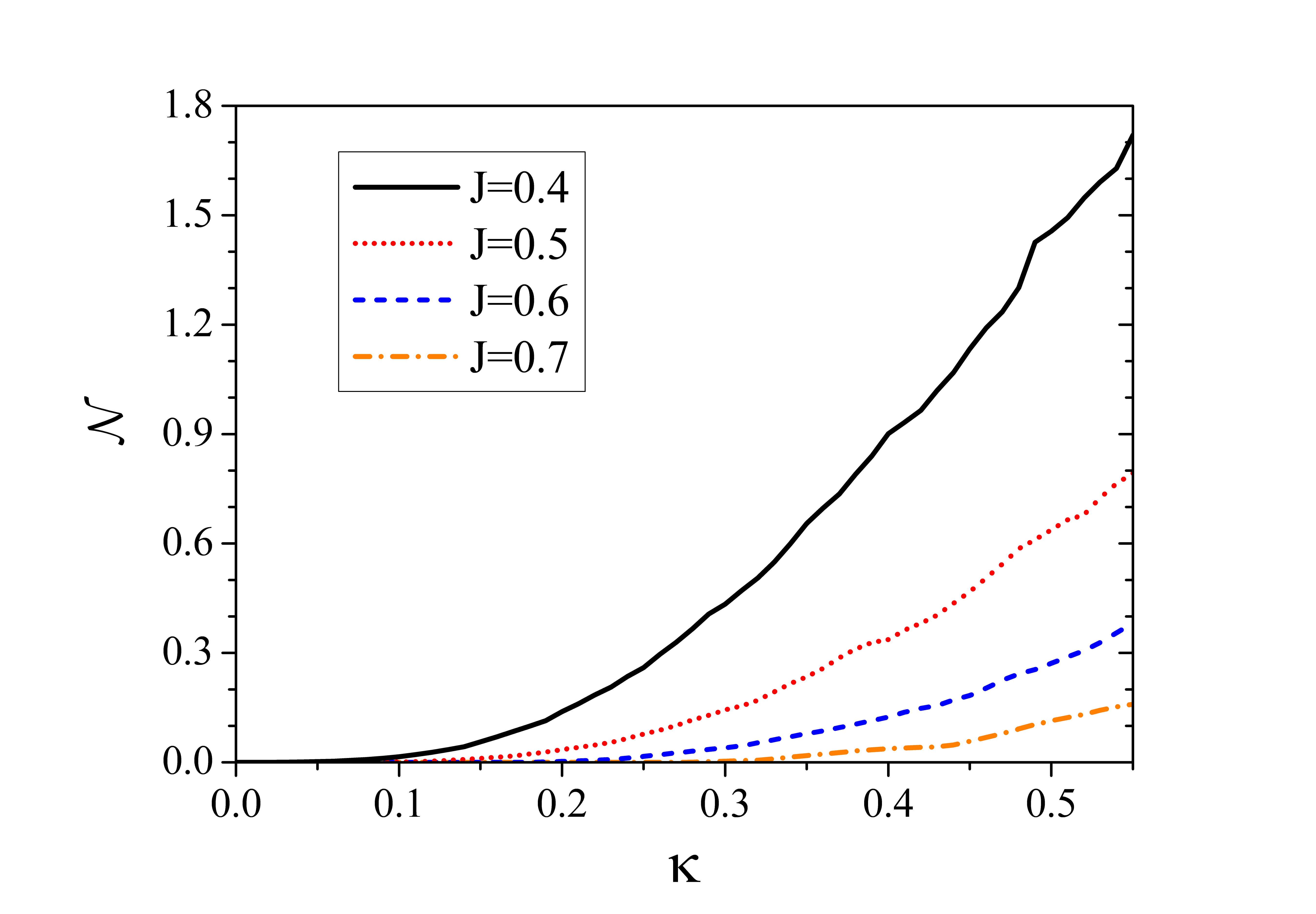}
}
\end{center}
\caption{(Color online) Non-Markovianity $\mathcal{N}$ versus $\kappa$ (the interaction strength between $S$ and $S'$) for different values of $J$ (the  interaction strength between $S'$ and environment qubits) in the absence of intra-interactions between environment qubits, that is $\Omega=0$. The remaining parameters are given as $T=1$ and $\omega=5\omega_0$.}
\label{M2NMkT1}
\end{figure}

In this subsection, we consider the non-Markovianity in the absence of collisions between environment qubits, i.e., for $\Omega=0$. For this indirect collision model, the information of the system $S$ is first transferred to the qubit $S'$ via the coherent interaction and then dissipated to the environment through the collisions between $S'$ and environment qubits.
In this case, the intermediate qubit $S'$ can have the role of a quantum memory leading to the non-Markovian dynamics even without collisions between environment qubits.
The interaction strength $\kappa$ between $S$ and $S'$ is then crucial in activating the non-Markovianity, as verified in Fig.~\ref{M2NMkT1} where $\mathcal{N}$ increases with $\kappa$ for a given $J$ at a fixed temperature. Moreover, the non-Markovianity achieves a nonzero value only when $\kappa$ is greater than a threshold and the larger the value of $J$, the larger the threshold of $\kappa$ required to trigger the non-Markovian regime. From Fig.~\ref{M2NMkT1} one also observes that, for a given $\kappa$, the non-Markovianity $\mathcal{N}$ decreases with $J$, which implies that a strong interaction between $S'$ and the environment qubits weakens the non-Markovianity of the system $S$.

\begin{figure}[t!]
\begin{center}
{\includegraphics[width=0.9\linewidth]{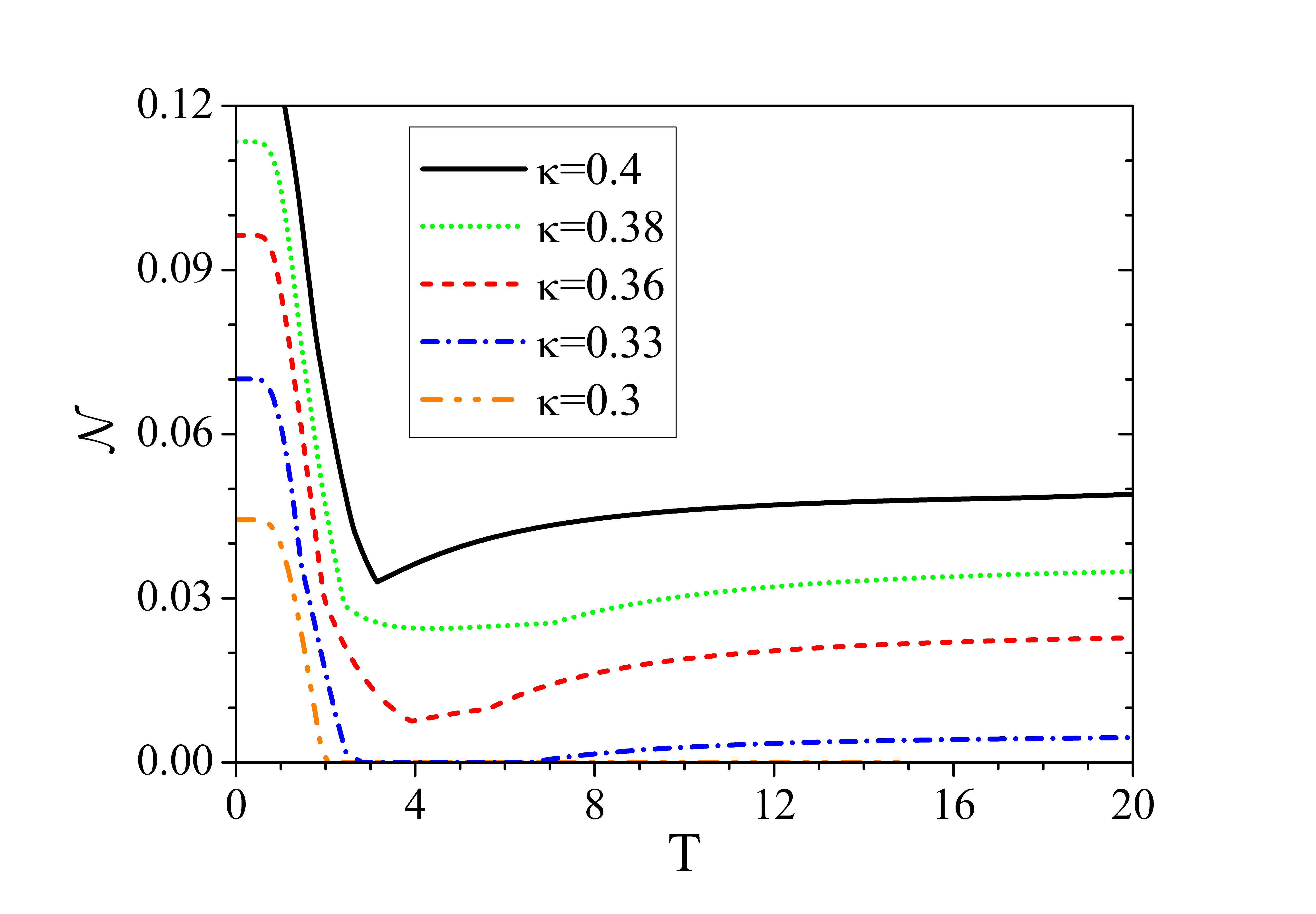} %
}
\end{center}
\caption{(Color online) Non-Markovianity $\mathcal{N}$ as a function of $T$ for different $\kappa$ with $\Omega=0$, $\omega=5\omega_0$ and $J=0.6=0.12\omega/\omega_0$.}
\label{NM-T-Omiga0}
\end{figure}

In Fig.~\ref{NM-T-Omiga0}, the effect of the temperature $T$ on the non-Markovianity is taken into account for different values of $\kappa$. We notice that the non-Markovianity as a function of $T$ is very sensitive to the value of $\kappa$, in that it can: decrease directly to zero (e.g., for $\kappa=0.30$), disappear for a finite range of temperature and then revive (e.g., for $\kappa=0.33$), or decrease to a minimum value and then slowly grow for larger $\kappa$. Remarkably, non-Markovianity can persist at high temperatures provided that the values of
$\kappa$ are sufficiently large.

\begin{figure}[t!]
\begin{center}
{\includegraphics[width=0.9\linewidth]{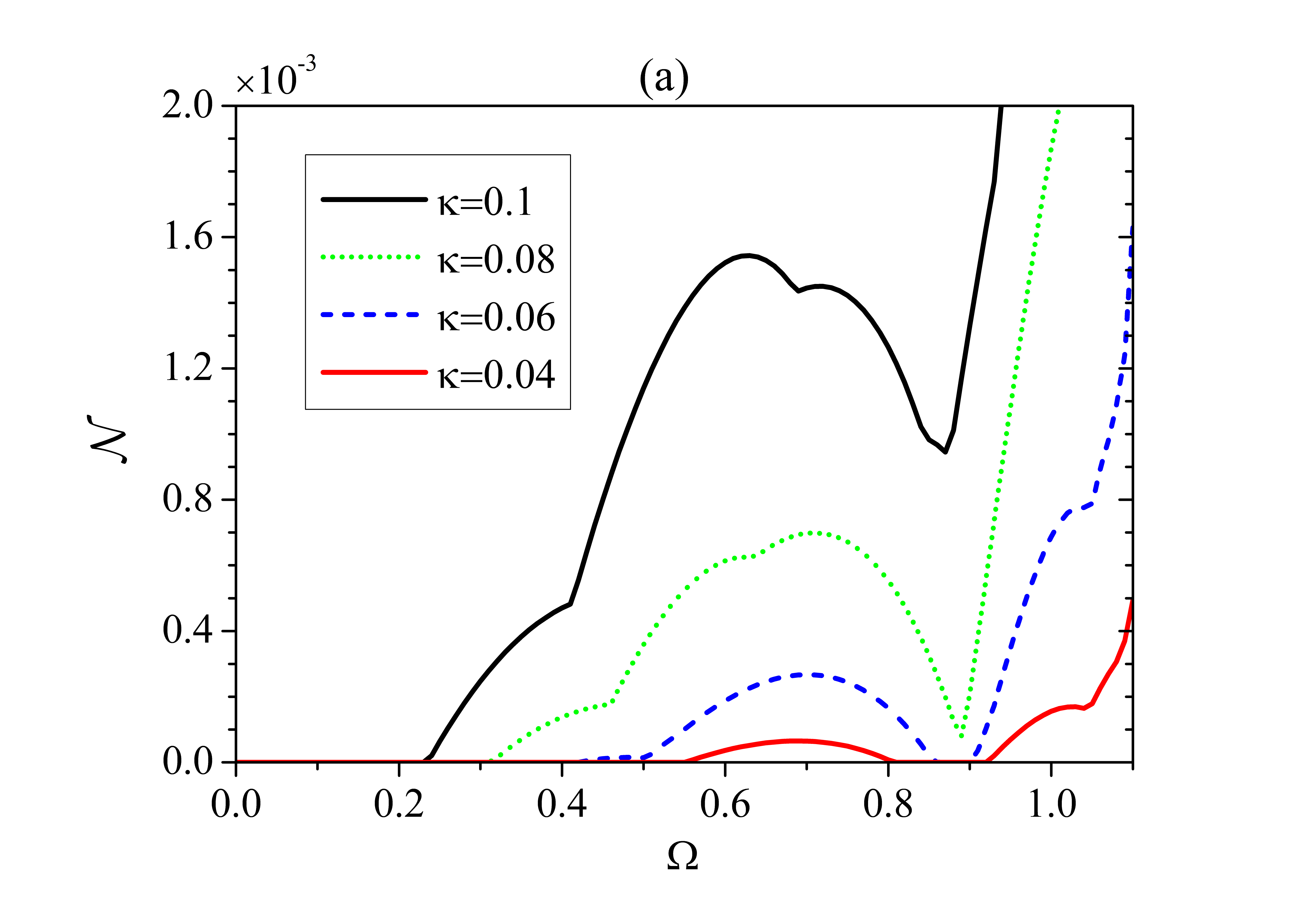}
\includegraphics[width=0.9\linewidth]{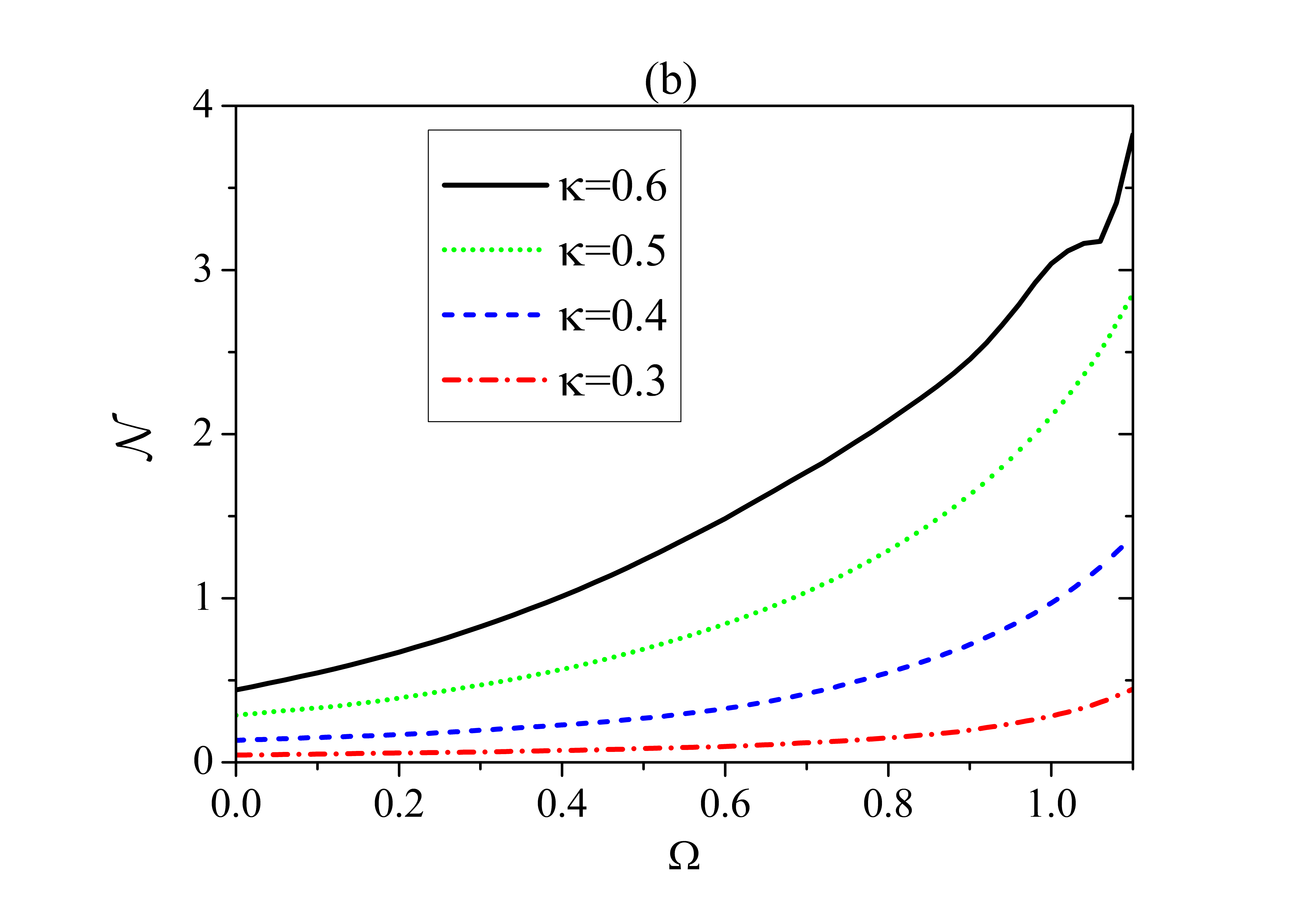}}
\end{center}
\caption{(Color online) Non-Markovianity $\mathcal{N}$ versus $\Omega$ for different $\kappa$ at $T=0$, with $\omega=5\omega_0$ and $J=0.6=0.12\omega/\omega_0$.}
\label{M2_NMOmiga_T0_J}
\end{figure}

\subsection{Presence of collisions between environment qubits}

\begin{figure}[t!]
\begin{center}
{\includegraphics[width=0.8\linewidth]{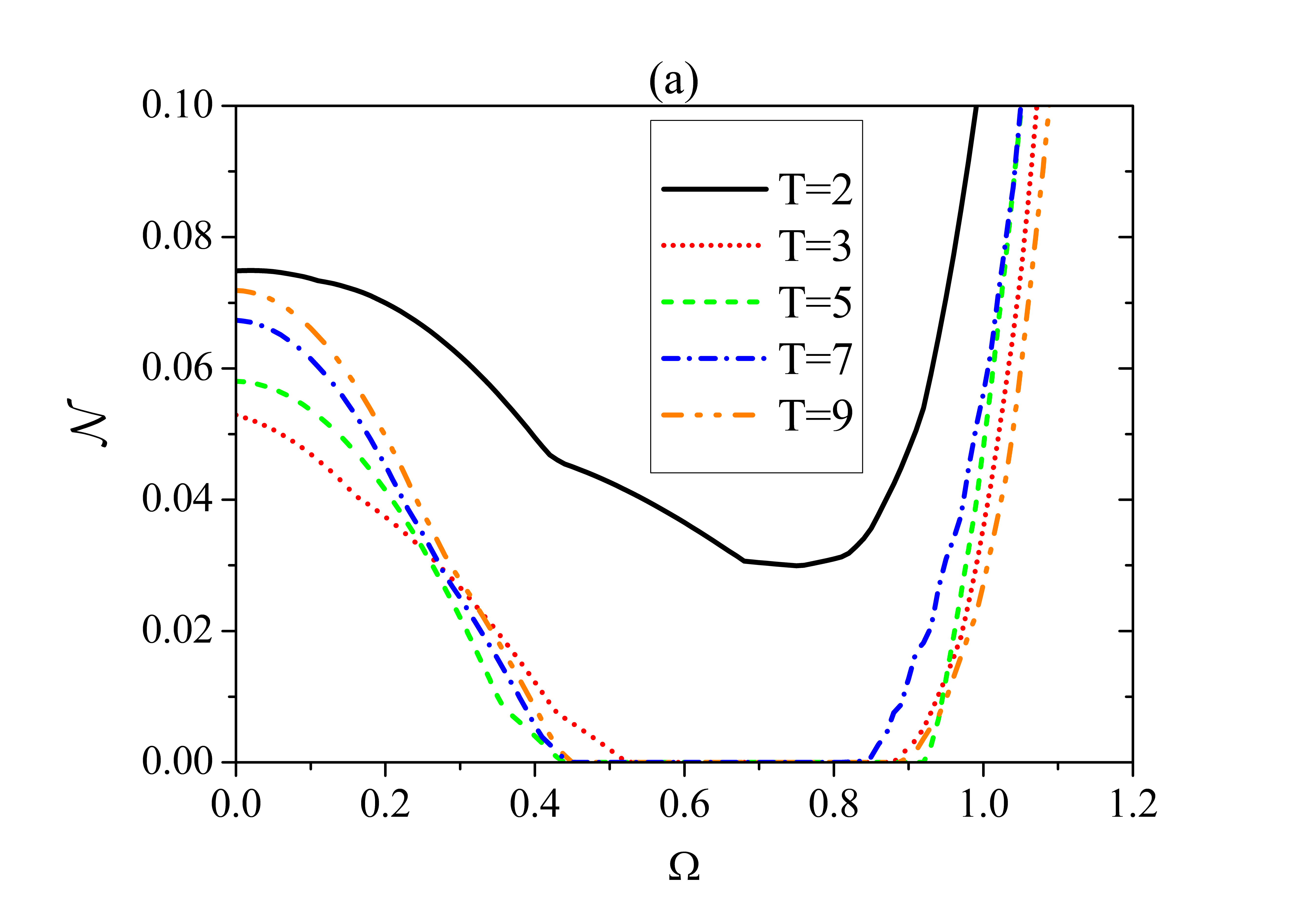}
\includegraphics[width=0.8\linewidth]{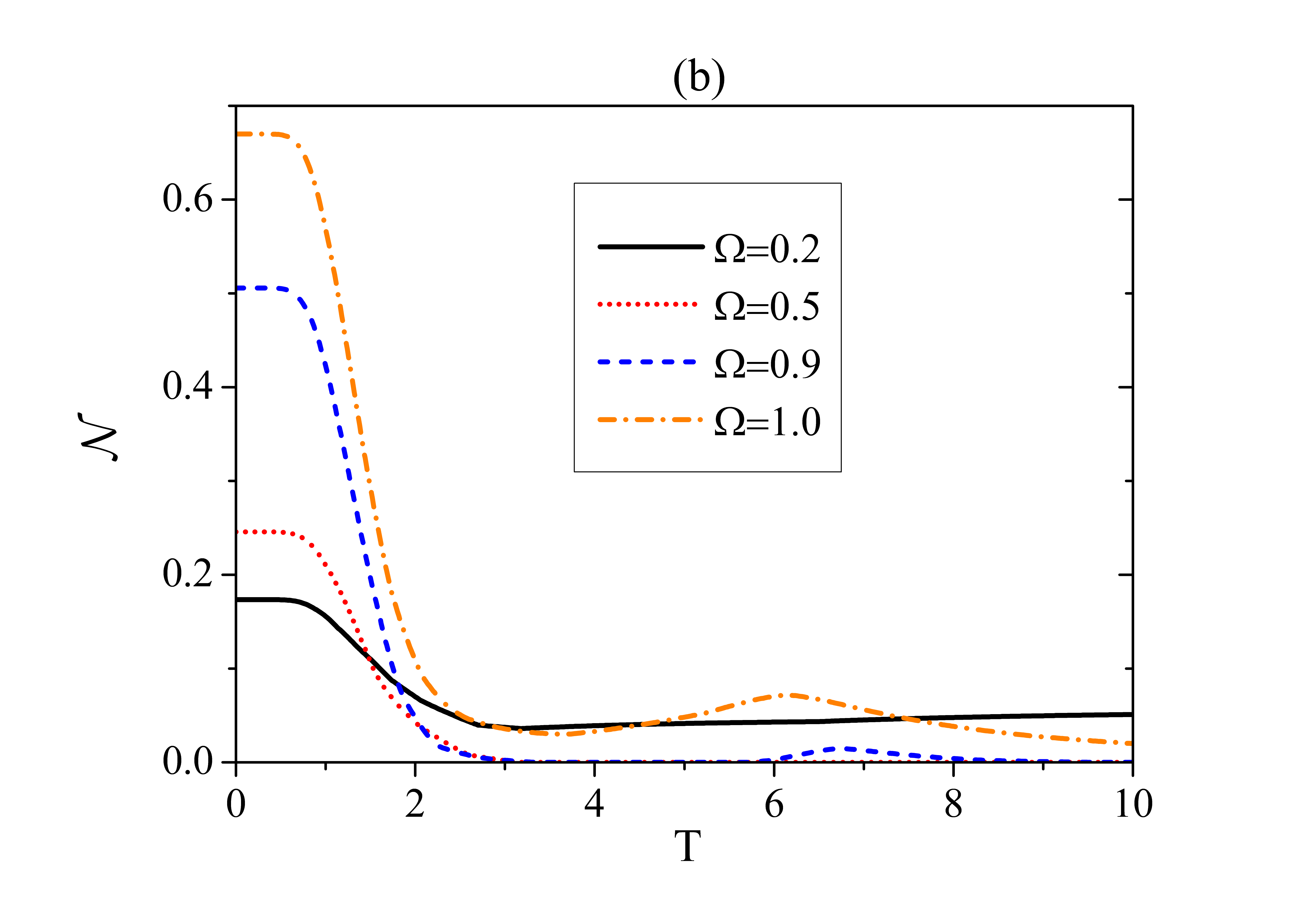}}
\end{center}
\caption{(Color online) (a) Non-Markovianity $\mathcal{N}$ versus $\Omega$ for different $T$. (b) Non-Markovianity $\mathcal{N}$ versus $T$ for different $\Omega$. The other parameters are given by $\omega=5\omega_0$, $\kappa=0.3$, $J=0.5=0.1\omega/\omega_0$.}
\label{M2_NMOmiga_T}
\end{figure}

Now we take the intracollisions between environment qubits $R_{n}$ and $R_{n+1}$ into account so that the two mechanisms of non-Markovian dynamics, namely the interaction
$S$-$S'$ ruled by $\kappa$ and the interaction $R_{n}$-$R_{n+1}$ ruled by $\Omega$, coexist in one and the same model.

We first explore the role of $\Omega$ in enhancing the non-Markovianity at zero temperature. When the coupling between $S$ and $S'$ is weak with relatively small $\kappa$, we know from the previous subsection that the dynamics of $S$ is Markovian ($\mathcal{N}=0$) if
$\Omega=0$. As shown in Fig.~\ref{M2_NMOmiga_T0_J}(a), by introducing the interactions $R_{n}$-$R_{n+1}$ a threshold of $\Omega$ exists which triggers a non-Markovian regime. Such a threshold increases with $\kappa$: namely, the smaller the value of $\kappa$, the larger the threshold of $\Omega$. The subsequent variations of $\mathcal{N}$ with $\Omega$ are non-monotonic. In particular, we find that the activated non-Markovianity $\mathcal{N}$ can disappear within a finite interval of $\Omega$ and then reappear (e.g., for $\kappa=0.04, 0.06$). When the coupling between $S$ and $S'$ is strong with larger $\kappa$, the dynamics may be already non-Markovian even for $\Omega=0$, as seen in Fig.~\ref{NM-T-Omiga0} at $T=0$ and shown more in detail in Fig.~\ref{M2_NMOmiga_T0_J}(b). In this case, the non-Markovianity can be further enriched by introducing the interactions between environmental qubits $R_{n}$-$R_{n+1}$.

\begin{figure}[t!]
\begin{center}
{\includegraphics[width=\linewidth]{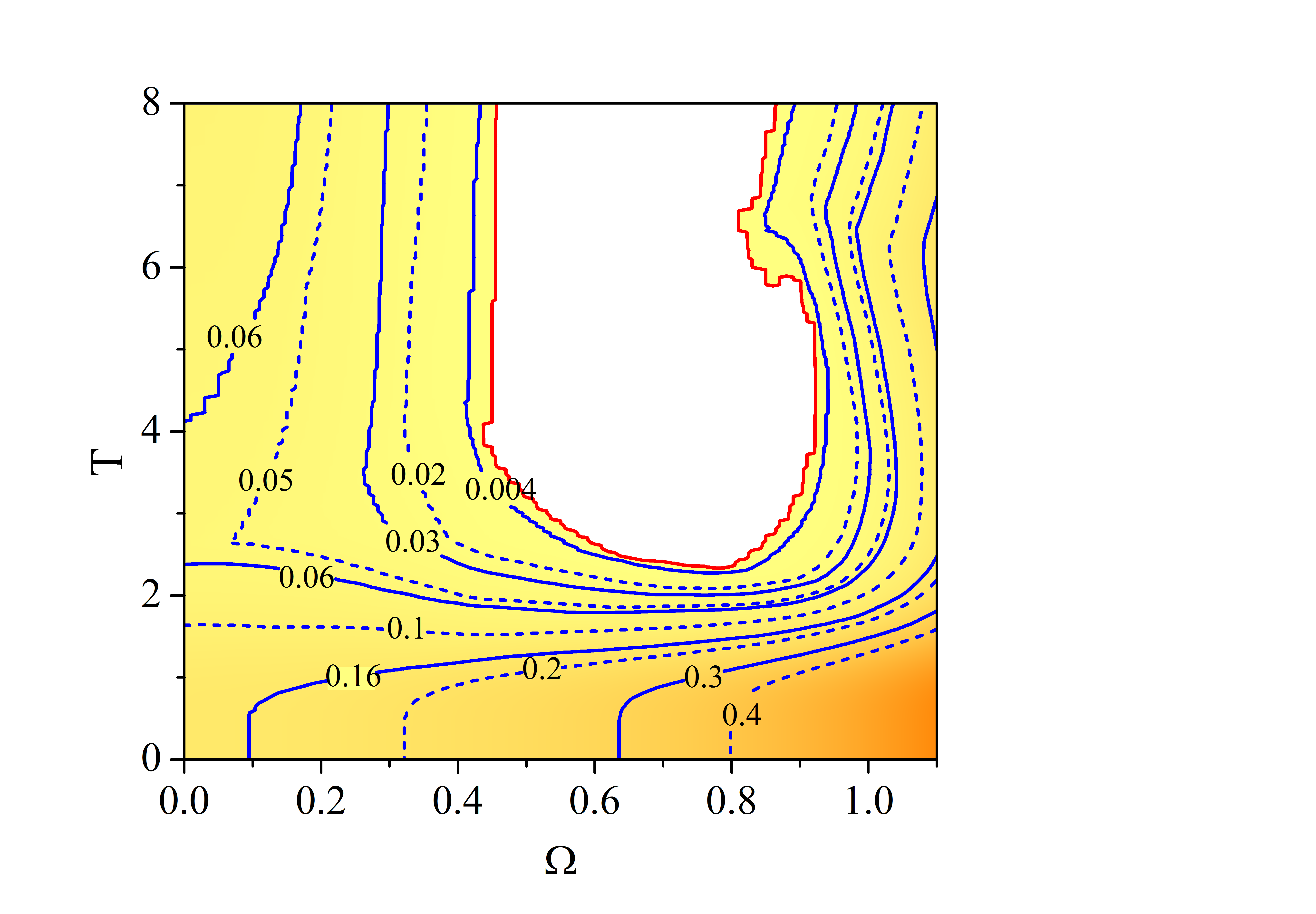}
}
\end{center}
\caption{(Color online) Contour plot of the non-Markovianity $\mathcal{N}$ for different $T$ and $\Omega$. The non-Markovian regime is colored while the Markovian regime is white. The alternate solid and dashed blue lines are the contour curves of $\mathcal{N}$. The
solid red line is the curve of the thresholds of $\Omega$ and $T$ triggering non-Markovianity. The other parameters are $\omega=5\omega_0$, $\kappa=0.3=0.06\omega/\omega_0$, $J=0.5=0.1\omega/\omega_0$.}
\label{CPM2NMTOmiga}
\end{figure}

The effects of the environment temperature on non-Markovianity are displayed in
Fig.~\ref{M2_NMOmiga_T}. In particular, $\mathcal{N}$ exhibits a non-monotonic variation with respect to the environmental qubit interaction strength $\Omega$ (see
Fig.~\ref{M2_NMOmiga_T}(a)), with a first descent and a successive ascent. Once again, we notice that the non-Markovianity can completely disappear for a finite range of $\Omega$ and then revive. The non-Markovianity $\mathcal{N}$ as a function of $T$ is then shown in
Fig.~\ref{M2_NMOmiga_T}(b), where we observe that the non-Markovianity is unavoidably weakened by increasing $T$ from zero for all the given $\Omega$, but it does not necessarily vanish for larger values of temperature. In fact, $\mathcal{N}$ can increase slowly (e.g., for $\Omega=0.2$), disappear completely (e.g., for $\Omega=0.5$), collapse and revive (e.g., for $\Omega=0.9$) and oscillate (e.g., for $\Omega=1.0$). A comprehensive picture of the variation of $\mathcal{N}$ as a function of both $\Omega$ and $T$, for fixed $\kappa$ and $J$, is given in Fig.~\ref{CPM2NMTOmiga}, where the above detailed behaviors can be retrieved. The values of $\kappa$ and $J$ are such that the system dynamics is non-Markovian for $T=0$ and $\Omega=0$. Such a plot is useful to immediately see how, in this composite indirect collision model, the temperature affects the system non-Markovianity in a different way from the case of the direct collision model treated in Sec.~\ref{DCM}. As a matter of fact,
Fig.~\ref{CPM2NMTOmiga} shows that the temperature has a general detrimental effect on non-Markovianity, which can never overcome its value at $T=0$ for the higher values of $T$, as instead happens for the direct collision model (see Fig.~\ref{NM-Omiga-T}(b)). However, a range of values of $\Omega$ exists for which a temperature threshold can be found which reactivate dynamical memory effects for the system lost at lower temperatures. In analogy with the direct collision model, such a feature is to be related to peculiar coherence exchanges from the system $S$ to the environmental components.

\section{Conclusion}

In conclusion, we have studied the effects of temperature on the non-Markovian character of an open quantum system dynamics by means of two types of collision models which entail different mechanisms for the occurrence of non-Markovianity.

In the first model, that is the direct collision model, the system $S$ consecutively interacts with a chain of environment qubits that are prepared in the same thermal states at temperature $T$, and the non-Markovianity $\mathcal{N}$ is induced by the intracollisions of environment qubits. As expected, the non-Markovian dynamics can be triggered when the intracollision strength is greater than a temperature-dependent threshold. In striking contrast to the usual understanding of the effect of the temperature on the non-Markovianity \cite{NMT1,NMT2,NMT3,NMT4,NMT5}, we have found that the behavior of $\mathcal{N}$ as a function of $T$ is non-monotonic, exhibiting a process of reduction and enhancement when temperature increases. In particular, we have shown that the non-Markovianity can vanish within a finite interval of $T$ and then reappear when $T$ increases. We have given a possible interpretation of this counterintuitive revival of dynamical memory effects by resorting to the exchanges of coherence between the system and environment qubits. In fact, albeit the temperature can accelerate the decay of coherence of the system and suppress the non-Markovianity until certain values, in the regime of high temperature this quick decay of system coherence can cause the coherence transferred to environment qubits to exceed that of the system. This mechanism in turn induces a backflow of information from the environment to the system and thus non-Markovian dynamics.

In the second model, that is the indirect collision model, the system $S$ indirectly interacts with the environment qubits through collisions with an intermediate qubit $S'$. In this case, $S'$  serves as the memory for the transferred information from $S$ towards the environment, representing a distinct non-Markovian mechanism. Without intracollisions between environment qubits, the non-Markovian dynamics for the system can still arise provided that the interaction strength of $S$-$S'$ is sufficiently large. Moreover, the non-monotonic relation between the non-Markovianity measure $\mathcal{N}$ and $T$ is once again observed. When the environmental intracollisions are taken into account, the two mentioned non-Markovian mechanisms coexist in the same model. In this case we have found that the presence of interactions between environmental qubits enriches non-Markovianity. The temperature has now the general effect to reduce the degree of non-Markovianity with respect to its value at zero temperature. However, once again non-Markovianity of the system can exhibit revivals as a function of the temperature.

Our findings within collision models are confirmed by some realistic composite quantum systems which exhibit a non-monotonic relation between non-Markovianity and temperature \cite{NMT4,ClosPRA}. More in general, our results contribute towards the capability of engineering suitable environments with optimal temperature conditions to exploit dynamical memory effects of an open quantum system, which is strategic for noisy intermediate scale quantum information processing \cite{NISQ}.

\acknowledgements
R.L.F. acknowledges Francesco Ciccarello for fruitful discussions and comments.
This work is supported by National Natural Science Foundation (China) under Grant Nos.~11574178 and 61675115, and Shandong Provincial Natural Science Foundation (China) under Grant No.~ZR2016JL005.


%

\end{document}